\documentstyle[12pt]{article}
\newcommand{\beq}{\begin{equation}}
\newcommand{\enq}{\end{equation}}
\begin{document}

\begin{titlepage}
\begin{flushright}
                Preprint KUL-TF-93/36 \\
                hepth@xxx/9308098 \\
                10.08.1993 \\
\end{flushright}
\vfill
\begin{center}
{\large\bf Chiral Splitting at Work.} \\
\vskip 27.mm
{\bf Z. Hasiewicz\footnote{Onderzoeker I.I.K.W. Belgium,
on leave from IFT University of Wroclaw, Poland}
, P. Siemion\footnote{On leave from IFT University
of Wroclaw, Poland. \\ Partially supported by KBN-grant 2 00 95 91 0 }}\\
\vskip 1cm
Instituut voor Theoretische Fysica
        \\Katholieke Universiteit Leuven
        \\Celestijnenlaan 200D
        \\B-3001 Leuven, Belgium\\[0.3cm]
\end{center}
\vfill
\begin{abstract}
Second-order equations of motion on a group manifold
that appear in a large class of so-called chiral theories are presented.
These equations are presented and explicitely solved for cases
of semi-simple, finite-dimensional Lie groups.
\end{abstract}
\vskip 27.mm
\thispagestyle{empty}

\end{titlepage}

\section{Introduction}
In this paper we would like to present a solution of a seemingly
complicated
differential equation by means of a trick known in 2-dimensional field
theory as
chiral splitting \cite{Wit}. The equations themselves arise in a quite
natural way
as equations of motion of a particle on a group manifold $G$. The phase
space is then the cotangent bundle $T^*G$, which can be identified with
$G\times{\cal G}$, provided there exists a non-degenerate $Ad$-invariant
quadratic form on ${\cal G}$ (Killing form). The group $G$ is equipped
with its natural left and right action on itself, but the liftings of
those
actions to actions on the bundle are not unique. The inequivalent
liftings
are labelled by group-cocycles $\theta$ described below.
Chosing particular liftings (and a particular cocycle)
typically requires a modification to the canonical symplectic structure
on $T^*G$ in order to make it invariant under the lifted actions.
The lifted actions are hamiltonian with respect to the modified
symplectic
form and they admit weak momentum mappings.
If the Hamiltonian is taken as a sum of the squares of the momentum
mappings
(for the canonical lift corresponding to $\theta = 0$ that gives
$H = {1\over 2}p^2$) there is a family of
dynamics corresponding to different group cocycles.
\par
Many details and much deeper discussion of this geometry can be found in
\cite{my} \cite{my2}. Here we would like to concentrate rather on the
equations themselves.
\section{The Equations}
Consider a phase space $G\times {\cal G}$, where $G$ is a Lie group
and ${\cal G}$ is its Lie algebra. Of course it can be
identified with $TG$. We assume that there exist an $Ad$-invariant
metric
$K$ on ${\cal G}$.
\par
Let $\theta$ be a group cocycle, i.e.
\beq
        \forall g_1,g_2 \in G \qquad \theta (g_1 g_2) = \theta (g_1)
        + Ad_{g_1} \theta (g_2) \quad \in {\cal G} ,
        \label{teta}
\enq
and let $\Sigma$ be the derivative of $\theta$:
\beq
        \Sigma (X) :=  {d \over dt} \theta (e^{tX}) |_{t=0}
        \quad \forall X \in {\cal G} .
        \label{sig}
\enq
We assume that $\theta$ is symplectic with respect to $K$, i.e.
\beq
        K(\Sigma(X)\,,\,Y) = - K(X\,,\,\Sigma(Y))
        \quad \forall X,Y \in {\cal G}
\enq
and the symplectic structure on $G \times {\cal G}$ is given by the
2-form:
\beq
        \Omega = d K(p,g^{-1}dg) - K(Ad_g^{-1} \theta(g), d g^{-1}dg) .
\enq
The moments corresponding to the right and left lifted actions are:
\beq
 J^r := p + \theta(g^{-1}) \quad ; \quad J^l := - Ad_g p - \theta (g) .
        \label{curr}
\enq
For thee Hamiltonian
\beq
        H={1\over 4} K(J^r\,,\,J^r) + {1\over 4} K(J^l\,,\,J^l)
\enq
one gets the following equations of motion
\beq
        g^{-1}{\dot g} = p
        \label{eqmot1}
\enq
\beq
        {\dot p} = - \Sigma (\theta (g^{-1})) + [ p , \theta (g^{-1})].
        \label{eqmot2}
\enq
The equations (\ref{eqmot1}),(\ref{eqmot2}) are not transparent and
even if one is able to write down a formal solution, it is in general
not obvious what the trajectories look like etc\footnote{Note however
that
for $\theta=0$ one recovers the motion along geodesics ($\dot p =0$).}.

\section{Abelian case}
There is one case when the equations of motion (\ref{eqmot1}),
(\ref{eqmot2}) are particularly simple and this is when $G$ is abelian
($G\simeq {\bf R}^n$).
Then the $Ad$ representation is trivial and the condition (\ref{teta})
tells us that $\theta$ is a linear operator. The derivative $\Sigma$ is
numerically equal to $\theta$ (although it acts on a different space),
the translations in velocities are trivial and (\ref{eqmot2}) reduce to
\footnote{We write $x$ instead of $g$ as it is an element of a vector
space
here.}
\beq
        \ddot x = - \theta^T \theta x .
        \label{abem}
\enq
Here $\theta^T$ is a transposition defined with respect to the Euclidean
form on ${\bf R}^n$.
The r.h.s of (\ref{abem}) is never positive.
Therefore in the abelian case the motion is either oscillatory
(negative eigenvectors) or free (null directions of $\theta$).

\section{Semi-simple case}
On a semi-simple, finite-dimensional Lie group $G$ the cocycle $\theta$
is neccessarily of the form:
\beq
        \theta (g) = Ad_g \mu  - \mu  ;
        \quad \mu  \in {\cal G}
        \label{trcoc}
\enq
Then from (\ref{sig}) one has:
\beq
        \Sigma(X) = ad_X \mu \quad \forall X \in {\cal G}
\enq
and the equations of motion take the form:
\beq
\matrix{ g^{-1}\dot g = p  ; &
        \dot p = [p + \mu, Ad_g^{-1}\mu - \mu] }.
        \label{emu}
\enq
Solving these equations seems difficult, because the 'brute force'
methods
are likely to give the solution in form of ordered exponents, which are
hardly transparent. Nevertheless it turns out that there is a simple
formula for
the solution of (\ref{emu}) for a starting point $(g_o,p_o)$, given by:
\beq
        g(t) =  exp\{-\mu t\}\,g_o\,exp\{(p_o+ \mu + Ad_{g_o} \mu)t\}\,
        exp\{-\mu t\}.
        \label{solmu}
\enq
Deriving (\ref{solmu}) relies heavily on parametrizing as much of the
phase space as possible by 'chiral momenta' (\ref{curr}), which satisfy
simple, time independen linear equations:
\beq
\matrix{  \dot J^l = \Sigma (J^l) ; & \dot J^r = - \Sigma (J^r) } .
\label{wzw}
\enq

\par
Let us assume the following Ansatz for the solution \footnote{
In fact this means considering the model as a symplectic reduction of a
larger system by suitable constrains. The details of this geometric
interpretation can be found in a forthcoming paper.}:
\beq
        g(t) = u_l(t)\, u_r(t).
        \label{product}
\enq
Then using (\ref{eqmot1}) and the cocycle property (\ref{teta}) we can
write
\beq
   J^r(g,p) = u_r^{-1} \dot u_r + \theta (u_r^{-1}) +
        Ad_{u^r}^{-1} \big (u_l^{-1} \dot u_l +
        \theta (u_l^{-1}) \big )
\enq
and
\beq
        - J^l (g,p) = \dot u_l u_l^{-1} + \theta (u_l) +
        Ad_{u_l}\big (\dot u_r u_r^{-1} +
        \theta (u_r)\big ).
\enq
Introducing new variables
\beq
        \xi_l := u_l^{-1} \dot u_l + \theta (u_l^{-1})
        \label{xil}
\enq
and
\beq
        \xi_r := \dot u_r u_r^{-1} + \theta (u_r)
        \label{xir}
\enq
we can write
\beq
        J^r = Ad_{u_r}^{-1} \big ( \xi_l + \xi_r - 2\theta (u_r) \big )
\enq
\beq
        - J^l = Ad_{u_l} \big (\xi_l+\xi_r-2\theta (u_l^{-1}) \big )
\enq
A straightforward calculation shows that each of the equations
(\ref{wzw})
is equivalent to:
\beq
        \dot{( \xi_l +  \xi_r)} + (\Sigma -{1\over 2} ad_{\xi_l+\xi_r}
        (\xi_l-\xi_r) = 0 .
        \label{strain}
\enq
This resembles the zero-curvature equation (the Gauss law constraint)
of chiral two-dimensional field theory \cite{Wit}.
The set of solutions to (\ref{strain}) is much too large for our
purposes.
We shall restrict $\xi_l$ and $\xi_r$ (simplifying the equations at
the same time) by using an obvious gauge invariance of (\ref{product}).
The transformation $u_r \rightarrow h u_r , u_l \rightarrow u_l h^{-1}$
with $h$ arbitrarilly time-dependent, leaves (\ref{curr}) and
(\ref{strain})
invariant, while (\ref{xil}),(\ref{xir}) change as follows:
\beq
        (\xi_l+\xi_r) \rightarrow Ad_h(\xi_l+\xi_r) + 2\theta(h)
\enq
\beq
        (\xi_l-\xi_r) \rightarrow Ad_h(\xi_l-\xi_r) - \dot h h^{-1}
\enq
{}From the above equations one can see that the dfference $\xi_l-\xi_r$
can always be transformed to be zero. Moreover, such a 'gauge' remains
invariant under time-independent transformations.
\par
Then the equation (\ref{strain}) tells us that $\xi = \xi_l= \xi_r$ is
time independent. Also, using time independent gauge transfofmations,
one can allways make $\xi$ belong to the kernel of $\Sigma$ .
\par
The equations (\ref{curr}) can be written as:
\beq
        p + Ad_{u_lu_r} \mu + \mu = 2 Ad_{u_r}^{-1}( \xi +\mu )   ,
        \label{incon1}
\enq

and the equations (\ref{xil}),(\ref{xir}) can be regarded now as simple
equations of motion for $u_r$ and $u_l$ respectively:
\beq
\matrix{ \dot u_l =  - \mu u_l + u_l (\mu + \xi_l )    \cr \cr
         \dot u_r  =  ( \xi_r + \mu )u_r - u_r \mu }
         \enq
with obvious solutions given by:
\beq
        u_l(t) = exp\big\{-\mu t\big\}\,u_l(0)\,
        exp\big\{(\mu+\xi_l)t\big\}
        \label{ult}
\enq
\beq
        u_r(t) = exp\big\{(\mu+\xi_r)t\big\}\, u_r(0)\,
        exp\big\{-\mu t\big\} .
        \label{urt}
\enq
Now taking $g(t) = u_l(t)u_r(t) $ gives
\beq
        g(t) = exp\big\{- \mu t\big\}\,u_l(0)\,
        exp\big\{(2\mu+\xi_l+\xi_r)t\big\}\, u_r(0)\,
        exp\big\{-\mu t\big\}
        \label{godt}
\enq
which is exactly equal to (\ref{solmu}) because from (\ref{incon1})
one has
\beq
\matrix{ u_l(0)\, exp\big\{ (2\mu+\xi_l+\xi_r) t\big \} u_r(0) =\cr\cr
        =  g(0) exp\{ (p(0) + \mu + Ad_{g(0)}\mu)t \}
        }.
\enq

\section{Su(2) (pictures)}
In this case of low dimensional group it is possible to plot some
of the trajectories.\\

In its fundamental representation Su(2) is a group of $2 \times 2$
matrices of the form:
\begin{equation}
\matrix{ g = \pmatrix{ a & - \bar b \cr b & \bar a } &\quad ;
        \quad  |a|^2 + |b|^2 = 1 }
        \label{gst}
\end{equation}
where $a,b \in {\bf C}$ and bar denotes
complex conjugation. The Lie algebra $su(2)$ is spanned by the
antihermitean generators $l_k ; k = 1,2,3 $ proportional to Pauli
matrices. The momentum is therefore given as
\beq
        p = p^k l_k ;\quad k=1,2,3.
\enq
Su(2) is a simple group and therefore every cocycle is of the
form (\ref{trcoc}) and without any loss of generality we can assume that
\beq
        \theta (g) = m (Ad_g l_3 - l_3) ;
        \label{thetal}
\enq
where $m \in {\bf R}$ is an arbitrary parameter and $l_3$ is an
anti-hermitian matrix:
\beq
l_3 = \pmatrix{ {i\over 2} & 0 \cr 0 & -{i\over 2} } .
\enq
The group is identyfied with $S^3$ and therefore general trajectories
on it may be difficult to present. Howewer, for trajectories starting
at the group unity (or any other element of $G$ of the form $e^{tl_3}$)
a straighforward calculation shows
that for evolution given by (\ref{solmu}) the phase of $b$ of (\ref{gst})
is constant. Therefore one is left with three real parameters:
$Re(a),Im(a),|b|$, and these
trajectories can be presented as lying on $S^2$. The figures (1,2,3)
present some of these trajectories plotted by means of the
{\it Matematica}$^{TM}$ program.
\par
Note that the model is in a sense 'linear', i.e. a shape of a trajectory
depends only on the ratios of $p_k$ and $\mu$ and mutiplying them
by a constant amounts just to rescaling of the time variable $t$.
The real parameter $m \in {\bf R}_+$ is in many respects similar to
a strenght of magnetic field.  In particular if initial momentum
is parallell to $l_3$ then the motion is 'free' (along the big circle).
\par
Figure 1. presents beginnings of a family of trajectories starting
at the group unity (the top of an imaginary ball) for fixed
$p = (0,-2,4)$
and different values of $m$. For $m =0$ the trajectory is a circle: As
we
have already pointed out if $m=0$ the trajectories are just geodesics.
As $m$
increases the circle gets more and more distorted. Then for $m=2.11$
it closes again and then opens again etc.
\par
Actually, it is enough to look at (\ref{solmu}) to see that it is
sufficient
to have $|p+2\mu|s = |\mu|r$ for some $s,r \in {\bf Z}$ to have a closed
trajectory, but the period may be relatively long.
\par
Figure 2. presents (a beginning of) a typical trajectory with initial
velocity perpendicular to $\mu$. The actual values are $p=(-1,0,0), m=-5$
and $t=15$.
\par
Figure 3. presents a trajectory for $p =(-1,2,0)$ and $m=4.085$. The
feature of it we would like to point out are 'loops'. As the value
of $m$ increases they are getting smaller and eventually vanish.
At this point a trajectory has 'cusps' i.e. there are points at
which the velocity drops to zero. This is very similar to a cycloidal
motion
of a particle in crossed electric and magnetic fields.

\end{document}